\documentclass[aps,prx,10pt,twocolumn,groupedaddress,nofootinbib,notitlepage,showpacs,floatfix,superscriptaddress,longbibliography]{revtex4-1}

\usepackage{graphicx,graphics,epsfig,subfigure,times,bm,bbm,amssymb,amsmath,amsfonts,amsthm,mathrsfs,MnSymbol}
\usepackage[matrix,frame,arrow]{xypic}
\usepackage[pdfstartview=FitH]{hyperref}
\usepackage{subfigure}
\hypersetup{
    colorlinks=true,       
    linkcolor=red,          
   citecolor=magenta,        
    filecolor=magenta,      
    urlcolor=cyan,           
    runcolor=cyan
}
\usepackage[pdftex]{color}
\usepackage{amsmath}
\usepackage{braket}
\usepackage{enumerate}
\usepackage[normalem]{ulem}
\usepackage[usenames,dvipsnames]{xcolor}
\usepackage{multirow}
\usepackage{mathtools}

\definecolor{orange}{rgb}{1,0.5,0}

\newcommand{\bes} {\begin{subequations}}
\newcommand{\ees} {\end{subequations}}
\newcommand{\bea} {\begin{eqnarray}}
\newcommand{\eea} {\end{eqnarray}}

\definecolor{gold}{rgb}{0.85,.66,0}

\newcommand{\beq}{\begin{equation}}

\newcommand{\eeq}{\end{equation}}

\newcommand{\ignore}[1]{}

\def\s{\sigma}

\def\>{\rangle}
\def\<{\langle}
\def\Tr{\mathrm{Tr}}

\def\s0{I}

\newcommand{\ig}[1]{}

\begin{document}
\title{Entanglement enabled telescopic arrays in the presence of decoherence}
\author{Siddhartha Santra, Brian T. Kirby, Vladimir S. Malinovsky, Michael Brodsky\\ \it{U.S. Army Research Laboratory, Adelphi, MD 20783}}

\begin{abstract}

We consider quantum enhancement of direct-detection interferometric measurements to increase telescope resolution. We propose a protocol of measuring interferometric visibility function using imperfectly entangled states shared between remote telescopes. We show how errors in visibility measurement, and in turn, errors in intensity distribution of a distant object depend on the degree of entanglement of the shared quantum resource. We determine that these errors are sufficiently small over a wide range of resource states which makes our technique feasible in practical environments.
 \end{abstract}
\maketitle

\section{Introduction}

The angular resolution of telescopic arrays used in direct-detection interferometric measurements can be enhanced by increasing the baseline size, the distance between telescopes. The observed interference patten contains information about the correlation function of the radiation from an astronomical object and allows the extraction of information about the amplitude and phase of the complex visibility function, often called the fringe parameter, mutual intensity, or mutual coherence function. Increasing the baseline of the telescope array while maintaining sensitivity can improve the resolution of the source intensity distribution~\cite{review-monnier}. However, one problem with the direct-detection interferometric method is the loss of photons during transmission between the telescopes in an array~\cite{review-monnier}. Longer baselines lead to higher photon loss resulting in lower rates of successful interference detection events, which in turn reduces the scheme sensitivity.

A way to mitigate this problem using mode-entangled photons has been proposed recently in~\cite{prl-gottesman-lb}.
The main idea was to distribute known and replaceable photons in a perfect Bell-state between two telescopes in advance, utilizing a quantum network~\cite{book-vanmeter,review-qn-duan}, and extract the visibility function from local measurements, therefore eliminating the propagation loss of the collected photons. However, the technology required to reliably distribute perfectly entangled quantum states such as high-throughput repeaters~\cite{repeater-dur,repeater-bri,repeater-lukin}, long-lifetime quantum memories~\cite{memory-review}, decoherence-free entanglement swapping mechanisms~\cite{pra-brian}, high-fidelity quantum gates for purification~\cite{pur1}, distillation~\cite{distillation} and error-correcting protocols~\cite{surfacecodes} is not mature enough to yield distributed states with fidelities close to perfectly-entangled Bell-states~\cite{gen-reps}.

In this work we consider quantum enhancement of the direct-detection interferometric measurements using realistic, imperfectly-entangled quantum states as a resource. We propose to measure the complex visibility function utilizing quantum $X$-states which could feasibly be distributed across the nodes of a quantum network with currently available technology. The $X$-state form of the density matrix is general enough to take into account decoherence and photon loss in the distribution process between a perfect Bell-pair source and two telescope sites, and as examples we examine the effects of amplitude-damping, dephasing and depolarization on the entanglement distribution process~\cite{Brodsky:11,antonelli2011sudden,shtaif2011nonlocal}. Further, we calculate the dependence of the measurement rate and visibility on the resource state matrix elements and find that the results of the visibility measurements are determined by the concurrence \cite{conc-wootters} and the sum of diagonal density matrix elements (weight) of the $X$-state in the mode-entangled subspace. We also show that the error in the visibility magnitude is inversely proportional to the product of the concurrence and the square-root of the weight, while the error in the visibility phase is inversely proportional to the square-root of the weight of the $X$-state.

The remainder of this paper is organized as follows.
Sec.~(\ref{TnA}) briefly reviews interferometric measurement of the complex visibility function and the quantum-enhancement of this measurement using ideal Bell-pairs as a resource. Sec.~\ref{sec:imperfect} discusses the density matrix representation of the entangled resource states, the experimental scheme and measurement rates, the expected errors in the visibility measurements, and examines all of these results for specific channel decoherence models. Sec.~\ref{sec:conc} provides final conclusions and future work directions.


\section{Quantum-enhanced interferometry with ideal resources; visibility and resolution}
\label{TnA}

\subsection{Interferometric resolution}

\begin{figure}
\centering
\includegraphics[width=\columnwidth]{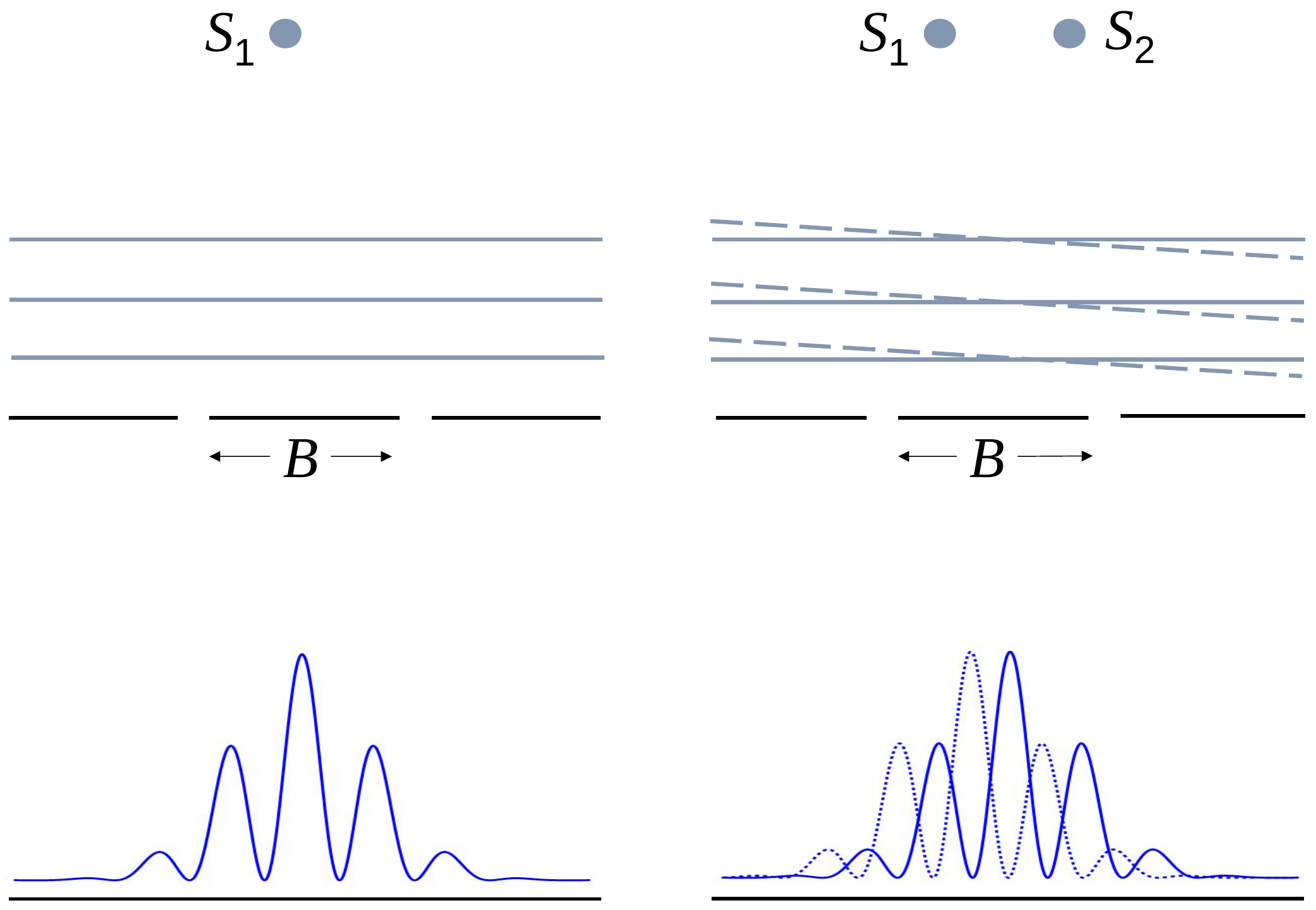}
\caption{(Color online) Left panel: Monochromatic plane waves of light from a distant source $S_1$ pass through two slits separated by a distance $B$. Constructive and destructive interference of the waves from the slits result in a pattern of alternating bright and dark fringes on a screen. Right panel: Similarly, two nearby sources $S_1$ and $S_2$ produce two sets of overlapping intereference patterns on the screen. These two sources are resolved if the maxima of one patten coincides exactly with the minima of the other.}
\label{fig1}
\end{figure}

Interferometric measurements allow us to extract the phase information of radiation collected from spatially separated points~\cite{review-monnier, lecnotes1}. This phase information can be used to distinguish the angular positions of different points at the source from which the radiation emerges, resulting in the resolution of different of different source points. The essential idea of interferometry can be understood through Young's double slit experiment, Fig.~(\ref{fig1}). Plane waves of monochromatic light from a distant point source interfere on a second screen upon passing through two slits (the distance between screens is negligible compared to the distance to the source) resulting in an intensity pattern of alternating bright and dark fringes. In the case of two point sources the interference patterns overlap. Two point-sources are resolved if the central maximum of the interference pattern from one source coincides with the first minimum of the interference pattern from the other. When this happens, the angular separation of the two point sources (resolution) is defined as
\begin{align}
\Delta\Theta=\lambda/2B,
\end{align}
where $\lambda$ is the wavelength of the monochromatic light, and $B$ the separation between the slits.

To determine the size of an extended source, the intensity distribution of the source as a function of the observation angle is measured. Usually, such sources are considered as multiple independent point-sources which produce correspondingly many overlapping fringe patterns in the interferometric measurement. The Van Cittert-Zernike theorem~\cite{zernike,vczbook} relates the contrast of the fringe pattern for an extended source, also called the visibility, to the Fourier transform of the source intensity distribution. We discuss the relationship between visibility and the source intensity distribution in the following section.

\subsection{Visibility}
\label{subsec:vis}

The complex visibility function $V_{\nu}(\bm r_1, \bm r_2)$ is defined as the spatial autocorrelation function of the electromagnetic radiation collected by telescopes at two different locations 
$\bm r_1$ and $\bm r_2$
\begin{align}
V_\nu(\bm r_1,\bm r_2):=\braket{\bar{E}_\nu(\bm r_1)\bar{E}_\nu^*(\bm r_2)},
\label{vis1}
\end{align}
where the raised asterisk denotes complex conjugation and subscript $\nu$ refers to the specific frequency for which the correlation is measured. Note, because $\bar{E}_1(\bm r_1)$ and $\bar{E}_2(\bm r_2)$ are 3D-vectors, Eq.~(\ref{vis1}) yields a tensor.
For simplicity, here we consider the electromagnetic field produced by the celestial sources to be a scalar. Thus we model the scalar field $\mathcal{E}(\bm R)$ produced at a distant point $\bm R$ in the sky propagating to the telescope at the observation point $\bm r$ via the propagator, $P(\bm R,\bm r)=e^{2\pi i\nu|\bm R - \bm r|/c}/|\bm R - \bm r|$, under the standard assumption of the space between the source and telescope being empty. The total field is given by adding up contributions from all sky regions on the celestial sphere with radius $|\bm R|$ which implies that at the point $\bm r_i$, the total field is $E(\bm r_i)=\int\mathcal{E}(\bm R)e^{2\pi i\nu|\bm R - \bm r_i|/c}/|\bm R -\bm r_i| d \bm \zeta$ with $d\bm \zeta$ an element of solid angle subtended by the source at the point $\bm r_i$. Under two other standard astronomical assumptions: a) the celestial sources are spatially incoherent, $\langle \mathcal{E}(\bm R_1)\mathcal{E}(\bm R_2)\rangle=0$, and b) far-field sources, $|\bm R_{1,2}| \gg |\bm r_{1,2}|$, Eq.~(\ref{vis1}) takes the form
\begin{align}
V_\nu(\bm r_1, \bm r_2)= \int I_\nu(\hat{s})e^{-2\pi i\nu \hat{s} (\bm r_1-\bm r_2)/c} d \bm\Omega ,
\label{vis2}
\end{align}
where $I_\nu(\hat{s})$ is the intensity distribution as a function of the observation direction vector $\hat{s}$, $d\bm\Omega$ is an element of solid angle, and integration is done over the entire solid angle subtended by the source.
Eq.~(\ref{vis2}) expresses the visibility function $V_\nu(\bar{r}_1,\bar{r}_2)$ as the Fourier transform of the source intensity distribution $I_\nu(\hat{s})$, at observation frequency $\nu$. Note that $I_\nu(\hat{s})$ is a function of the observation angle, given by the unit direction vector $\hat{s}$ relative to the fixed coordinate system at the telescopes. This is the essential content of the Van Cittert-Zernike theorem~\cite{zernike}: access to visibility at various baselines allows for complete reconstruction of the source intensity distribution. However, the range of the baseline-size is limited in experimental measurements, therefore using the inverse Fourier-transform of Eq.~(\ref{vis2}) we have
\begin{align}
I_\nu(\hat{s})&=\int_{0}^{\bm B_m} V_\nu(\bm r_1, \bm r_2) e^{2i\pi\nu\hat{s} \bm r/c} d \bm r,
\label{intdis}
\end{align}
where $\bm r =\bm r_2- \bm r_1$, $\bm B_m$ is the maximum baseline size.

\begin{figure}
\centering
\includegraphics[width=\columnwidth]{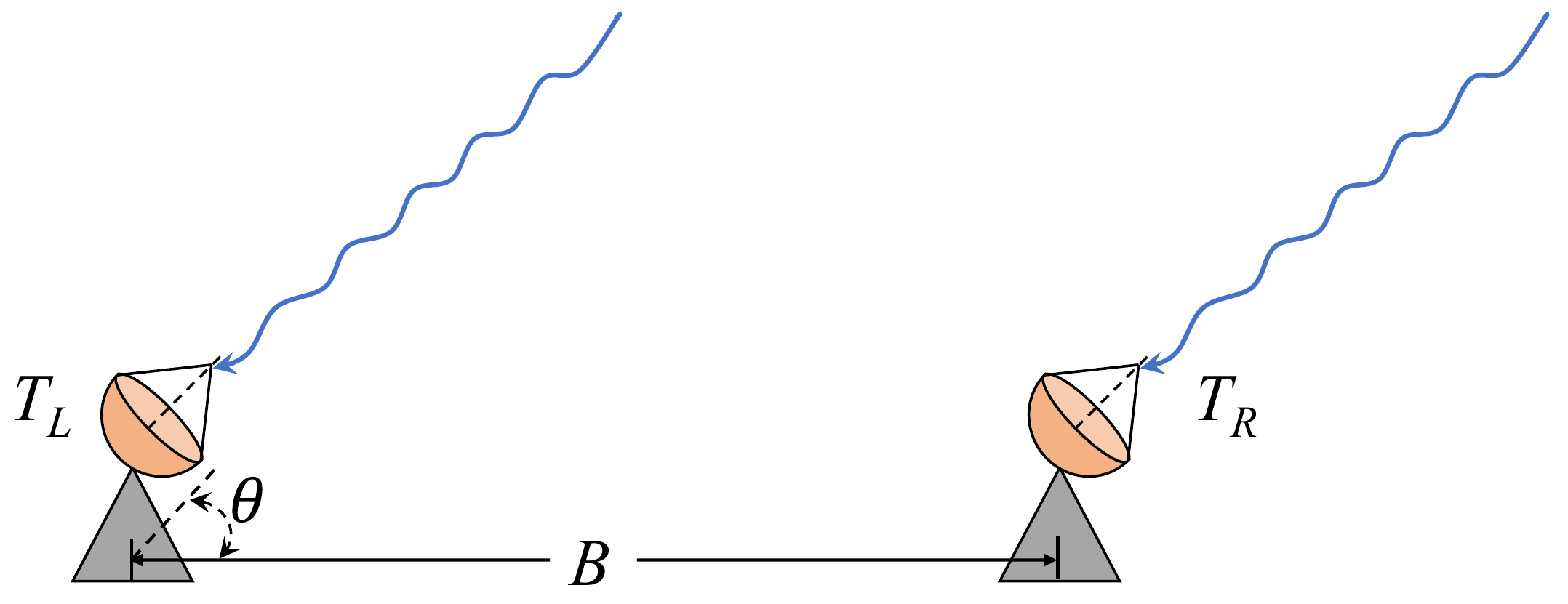}
\caption{(Color online) A pair of telescopes $T_L$ and $T_R$ separated by a distance of magnitude $B$ collect photons originating from distant sources. The collected photons have a phase difference proportional to the geometrical path difference given by $\Delta\phi=B \cos(\theta)/\lambda$.}
\label{Fig2}
\end{figure}

The relation in Eq.~(\ref{vis2}) implies that the angular-resolution of the interferometric array of telescopes can be improved by increasing the value of $\bm B_m$. Since direct-detection interferometric measurements require light to be physically brought from the telescopes to the cental detection station, where the correlation function in the R.H.S. of Eq.~(\ref{vis1}) is measured, the photon loss in optical channels limits the increase of the baseline size. In addition to photon losses, atmospheric density fluctuations and various physical mechanisms of noise also hinder the visibility measurements, reducing sensitivity and resolution of the direct-detection method~\cite{review-monnier,lecnotes1}. In this work, however, we only discuss how quantum-enhanced interferometry can mitigate the problem of photon-loss.

\subsection{Quantum-enhanced interferometry with ideal resources}
\label{subsec:lb}

\begin{figure}
\centering
\includegraphics[width=\columnwidth]{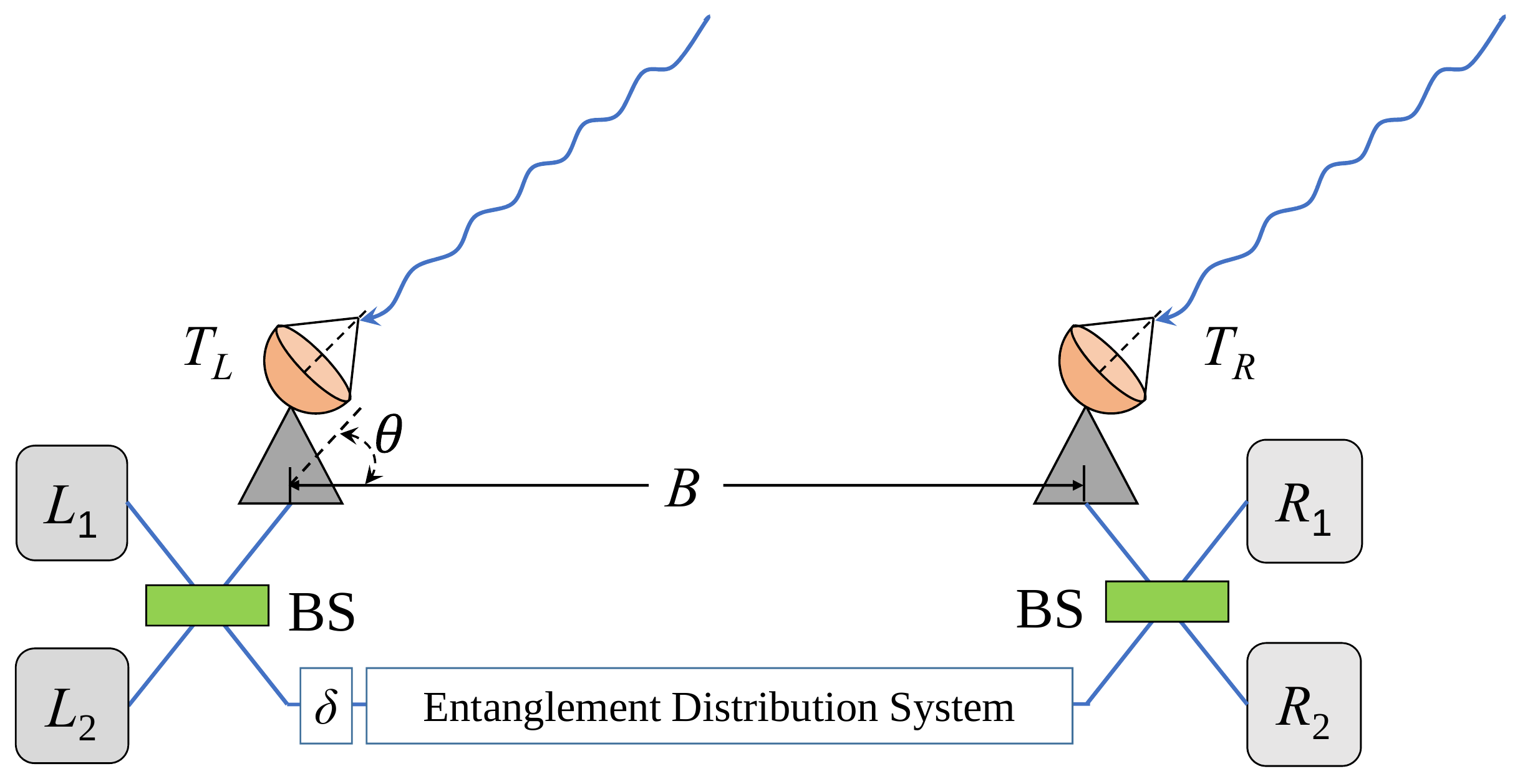}
\caption{(Color online) In quantum-enhanced interferometry the pair of telescopes $T_L$, $T_R$ separated by a baseline distance of magnitude $B$ share an entangled state. Correlations between outcomes of local measurements  performed on the incoming single photon from the target and the entangled photon supplied by the quantum network yield the desired interferometric information, i.e., visibility function. $\delta$ is a controllable phase difference between the left and the right paths.}
\label{Fig3}
\end{figure}

The essential idea of quantum-enhanced interferometry is to remove the need for actual physical transport of the collected photons from the separated telescopes to the measurement station. Instead, correlations between local measurements (at the telescope locations) on the collected photons from the astronomic object and an entangled pre-shared photons are used to obtain the visibility function.
The idealized scheme based on perfect Bell-states shared between the telescope sites is summarized in what follows. Gottesman et. al. \cite{prl-gottesman-lb} considered weak light from a distant astronomic object (denoted by the subscript $A$) at the single-photon level which is characterized  by a mode-entangled wave-function
\begin{align}
\ket{\psi}_A=\frac{1}{\sqrt{2}} \left(\ket{0}_L\ket{1}_R+e^{-i\Delta\phi}\ket{1}_L\ket{0}_R\right) ,
\end{align}
where $L,R$ denote the left and right telescopes, see Fig.~(\ref{Fig3}). Such astronomical single photons are rare, thus losses and noise incurred during the physical transmission process between the telescopes reduce the effective signal quality. Use of shared entangled states was proposed to overcome this problem of the signal degradation. A known perfectly-entangled photonic state is established between the telescope sites, in advance, before local joint-measurements on the astronomical photon and the network supplied photon are performed. It turns out that the correlations between these local measurements yield the same information of the visibility function as do direct interferometric measurements on the astronomic photons physically transmitted to the central detection station.

Let us give more details on the quantum-enhanced interferometry by considering an extended source comprised of multiple spatially-incoherent point sources. The ensemble of single astronomical photons from such an extended source is a probabilistic mixture of pure states of the form $\ket{\psi_A}$,  each with its own independent phase $\Delta\phi$. The general density matrix of the photons can be written in the form
\begin{equation}
\bm \rho_{A}=\frac{1}{2}\left(
\begin{array}{cccc}
 0 & 0 & 0 & 0 \\
 0 & 1 & \eta_{a}  e^{i \eta_{p}} & 0 \\
 0 & \eta_{a} e^{-i \eta_{p}} & 1  & 0 \\
 0 & 0 & 0 & 0 \\
\end{array}
\right) \, ,
\label{astrostate1new}
\end{equation}
where $\eta_{a}$ and $\eta_{p}$  are the real variables.

It is constructive to provide some details of full quantum mechanical description of the visibility function.
The direct-detection interferometry provides information about complex visibility, $V_a e^{i V_{p}}$, which is the first-order field correlation function~\cite{Berman,Scully}. For the pure single-mode quantum state of light, the field correlation function is defined as
\begin{align}
{\cal V}(\bm r_1, \bm r_2) = \bra{\psi_A}\bm E_L^{-}(\bm r_1) \bm E_R^{+}(\bm r_2) \ket{\psi_A} \, ,
\end{align}
where $\bm E_R^{+}\sim i a_R$ and $ \bm E_L^{-}\sim -i a^\dagger_L$ are the positive and negative frequency components of the electric field operator at the right and left telescopes (for simplicity spatial dependence is omitted here), $a^\dagger_R$ is the electromagnetic mode creation operator, and $\ket{\psi_A}$ is the state of the astronomical photons which can be expressed as
\begin{align}
\ket{\psi_A} =\frac{1}{\sqrt{2}}(\hat{a}^\dagger_R+e^{-i\Delta\phi}\hat{a}^\dagger_L)\ket{0_L0_R}\, .
\end{align}
After some algebra we find ${\cal V}(\bm r_1, \bm r_2) = e^{i\Delta\phi}/2$ , that means the visibility functions is defined by the relative phase between the modes of the single photon. For the extended source, using the photon density matrix in Eq.~(\ref{astrostate1new}) we have
\begin{align}
{\cal V}(\bm r_1, \bm r_2) = \Tr\{\bm \rho_A \bm E_L^{-}(\bm r_1) \bm E_R^{+}(\bm r_2) \} = \frac{1}{2} \eta_a e^{i \eta_{p}}\, .
\end{align}

As we see the visibility function is identical to the off-diagonal element of the density matrix in Eq.~(\ref{astrostate1new}), $\eta_a\equiv V_a$ and $\eta_p\equiv V_p$. Since the density matrix of the astronomical photon contains all information about the visibility function the general density matrix of the photon can be written in the form
\begin{equation}
\rho_{A}=\frac{1}{2}\left(
\begin{array}{cccc}
 0 & 0 & 0 & 0 \\
 0 & 1 & V_{a}  e^{i V_{p}} & 0 \\
 0 & V_{a} e^{-i V_{p}} & 1  & 0 \\
 0 & 0 & 0 & 0 \\
\end{array}
\right) \, ,
\label{astrostate1}
\end{equation}
which describes a mixed state for $|V_a|<1$ and a pure state if and only if $|V_a|=1$.

The authors in~\cite{prl-gottesman-lb} assumed that the shared entangled state between the telescopes is a pure mode-entangled Bell state. The state density matrix is given by
\begin{equation}
\rho_{E}=\frac{1}{2}\left(
\begin{array}{cccc}
 0 & 0 & 0 & 0 \\
 0 & 1 &  e^{-i \delta} & 0 \\
 0 & e^{i \delta} & 1  & 0 \\
 0 & 0 & 0 & 0 \\
\end{array}
\right) \, ,
\end{equation}
with $\delta$ the controllable phase-difference, between the paths to the left telescope and the right telescope from the photon source.
It was shown that the visibility function can be extracted through correlations between local measurements at the telescope sites. The setups consist of a $50:50$ beam-splitter (BS) with detectors at both output ports ($L1,L2$ at the left telescope and $R1,R2$ at the right telescope), Fig.~(\ref{Fig3}). If one uses the states $\bm \rho_A$ and $\bm \rho_E$  given above then the visibility can be obtained as correlations between detector clicks at the left and right telescopes; the probability of correlated ($L1,R1$) or ($L2,R2$) detections $p_{\text{c}}$ and that of anti-correlated ($L1,R2$) or ($L2,R1$) detections $p_{\text{ac}}$ are
\begin{align}
p_{\text{c}}&=\frac{1}{2} (1-V_{a} \cos (V_{p}-\delta)) \, ,\nonumber\\
p_{\text{ac}}&=\frac{1}{2} (1+V_{a} \cos (V_{p}-\delta)) \, .
\label{probs}
\end{align}
Thus, the amplitude, $V_a$, and phase, $V_p$, of the complex visibility can be deduced by scanning over different values of the controllable phase difference, $\delta$. By measuring the visibility for different baselines one can use Eq.~(\ref{intdis}) to eventually obtain the intensity distribution of radiation coming from the target.

\section{Interferometry with non-ideal resources}
\label{sec:imperfect}

In this section we discuss measurement of the visibility function using imperfectly-entangled quantum states as a resource. It is a tall order to require high-quality hardware that can provide long lived quantum memories, high-fidelity quantum gates, low-loss optical fibers, and other high-efficiency optical elements to distribute and store perfect Bell-pairs between two widely separated telescopes. In practice, decoherence in the quantum channel elements utilized for entanglement distribution leads to deterioration of the quantum resource quality ~\cite{sang-repeater}. For simplicity, here we refer to any entanglement distribution system as a quantum network. In the Subsec.~(\ref{subsec:decoh}) we consider several decoherence mechanisms in a  quantum network connecting the photon-source to the telescopes. We show that if the left and right arms of the network  are susceptible to independent channels of decoherence then it leads to shared entangled two-qubit states of the X-form. Here two-qubit refers to the 4-dimensional Hilbert space of a two-mode entangled single photon state. Then, in Subsec.~(\ref{subsec:vis-x}), we analyze the visibility function when the quantum state shared between the telescopes is an X-state, assuming perfect characterization of the shared entangled state. In practical scenarios the resource-state characterization may itself have errors~\cite{tom-error} which can also be taken into account by the explicit formulae for visibility function in terms of the resource-state density matrix elements as shown in Subsec.~(\ref{subsec:error}). However, in this work we only analyze errors in visibility measurements due to the non-perfect entangled nature of the resource state.

\subsection{Decoherence in the entanglement distribution channel}
\label{subsec:decoh}

\begin{figure}
\centering
\includegraphics[width=0.7\columnwidth]{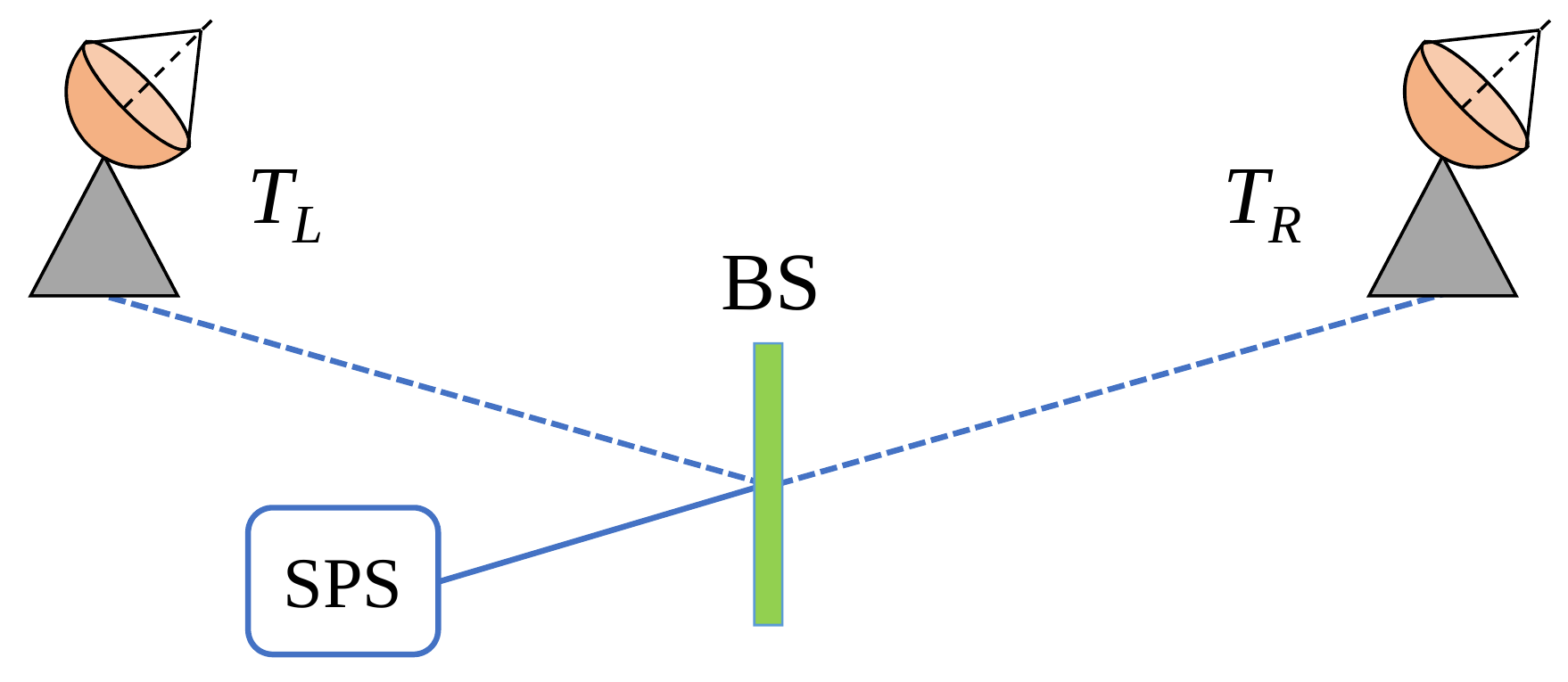}
\caption{(Color online) A simple scheme for entangled state distribution. The single photon source (SPS) and the $50:50$ beam-splitter (BS) create spatial mode-entangled Bell-pairs which are transmitted to the telescopes $T_L$ and $T_R$ via a quantum network. Depending on the total length of the quantum network, various methods of entanglement distribution, e.g. quantum repeaters, can be used.}
\label{Fig4}
\end{figure}

Various decoherence mechanisms in the quantum network lead to imperfectly entangled states. Independent decoherence in the two arms of the network connecting the telescopes, shown schematically in Fig.~(\ref{Fig4}), including the various connecting elements, quantum gates, filters etc., may generally be modeled via the amplitude damping, dephasing and depolarizing quantum channels. It can be shown (see Appendix~\ref{app:xform}), that if the entangled state generated by the single-photon source is one of the pure Bell-states (whose density matrix itself is of the X-form - all entries other than the main and anti-diagonal being zero) then any of the above decoherence mechanisms preserves the X-form for the resource state eventually received at the telescopes.

Derivation of the explicit forms of the resource state density matrices for the different decoherence mechanisms are in Appendix~\ref{app:xform}. Introducing the probabilities of the photon suffering amplitude damping  $\lambda_{L,R}$, dephasing $\mu_{L,R}$ and depolarization $\kappa_{L,R}$ in the left and right channels we obtain the following final shared resource states between the telescopes in the three cases,
\begin{equation}
\rho_{X}^{AD}=\frac{1}{2}\left(
\begin{array}{cccc}
 \lambda^{+}& 0 & 0 & 0 \\
 0 & \lambda_R^{-} & \sqrt{\lambda_L^{-}\lambda_R^{-} }& 0 \\
 0 &  \sqrt{\lambda_L^{-}\lambda_R^{-} }& \lambda_L^{-} & 0 \\
0 & 0 & 0 & 0 \\
\end{array}
\right)\, ,
\label{ampstate}
\end{equation}

\begin{equation}
\rho_{X}^{Deph}=\frac{1}{2}\left(
\begin{array}{cccc}
0 & 0 & 0 & 0 \\
 0 & 1 & \mu_L^{-}\mu_R^{-} & 0 \\
 0 & \mu_L^{-}\mu_R^{-} & 1 & 0 \\
0 & 0 & 0 & 0 \\
\end{array}
\right)\, ,
\label{dephstate}
\end{equation}

\begin{equation}
\rho_X^{Depol}=\left(
\begin{array}{cccc}
 x & 0 & 0 & 0 \\
 0 & 1/2-x & 1/2-2x & 0 \\
 0 & 1/2-2x & 1/2-x & 0 \\
 0  & 0 & 0 & x \\
\end{array}
\right)\, ,
\label{depolstate}
\end{equation}
where $\lambda^{+}=\lambda_R+\lambda_R$, $\lambda_{L,R}^{-}=1-\lambda_{L,R}$, $\mu_{L,R}^{-}=1-\mu_{L,R}$, $x=(\kappa_L+\kappa_R)/3-4\kappa_L\kappa_R/9$, $0\leq x\leq 1/3$, since $0\leq \kappa_{L,R} \leq 1$. Note that the $X$-form of the resource state is preserved even when the two arms undergo different decoherence mechanisms or a combination thereof as long as the decoherence in the two paths is independent.

\subsection{Visibility function measurements  using $X$-states}
\label{subsec:vis-x}

Using pre-shared entangled $X$-states, the visibility function can be obtained from local measurements of the probability for correlated and anti-correlated photon detections at the telescope sites, Fig.~(\ref{Fig3}), similar to the perfect Bell-pair case considered previously. The measurement setup consists of a $50/50$ beam-splitter at each telescope location with two single photon detectors at its outputs as shown in Fig.~(\ref{Fig3}). One input of each of the beam-splitters is from the respective telescope while the other input is for the network supplied photon. The two inputs to each of the beam-splitters thus correspond to a spatial mode of the astronomical photon and a spatial mode of the network photon. Each of the two single photons, one from the astronomical source and one from the network, may independently arrive at the two telescopes. The instances where these two photons arrive at different telescopes provide data useful for determining the visibility function.

We describe photon detections at the beam-splitter outputs by projection operators given by
\begin{align}
\Pi_{\pm}^{L}&=\ket{\psi_{\pm}^{L}}\bra{\psi_{\pm}^{L}}, \, \ket{\psi_{\pm}^{L}}=(\ket{1_A^L0_X^L}\pm\ket{0_A^L1_X^L})/\sqrt{2} \, ,\nonumber\\
\Pi_{\pm}^{R}&=\ket{\psi_{\pm}^{R}}\bra{\psi_{\pm}^{R}}, \, \ket{\psi_{\pm}^{R}}=(\ket{1_A^R0_X^R}\pm\ket{0_A^R1_X^R})/\sqrt{2} \, ,
\end{align}
where $L, R$ denote whether the measurement is at the left or right telescope,
the subscripts $A, X$ indicating an astronomical photon or the resource entangled photon. Correlated measurements at the right and left telescopes are identical measurements of either $\Pi_+$ or $\Pi_-$ at both telescopes, in this case pairs of detectors $L_1,R_1$ or $L_2,R_2$ detect the photons.

The probability of correlated photon-detections is given by
\begin{align} q_c=\Tr[\Pi_+^L\Pi_+^R\rho_A\otimes\rho_X]+\Tr[\Pi_-^L\Pi_-^R\rho_A\otimes\rho_X],
\end{align}
where $\rho_A$ is the density matrix of the astronomical single-photon defined in  Eq.~(\ref{astrostate1}),
and the shared entangled state between the two telescopes is now described by
\begin{equation}
\rho_{X}=\left(
\begin{array}{cccc}
 a & 0 & 0 & e^{-i z_{p}} z_{a} \\
 0 & g & e^{-i w_{p}} w_{a} & 0 \\
 0 & e^{i w_{p}} w_{a} & f & 0 \\
 e^{i z_{p}} z_{a} & 0 & 0 & h \\
\end{array}
\right) \, ,
\label{xstate}
\end{equation}
where $a+g+f+h=1$. Note, that $\rho_X$ encompasses states of the form in Eqs.~(\ref{ampstate}),(\ref{dephstate}), (\ref{depolstate}), i.e., states arising out of all possible modes of decoherence which occur independently in the two arms of any entangled-state distribution-system such as shown in Fig.~(\ref{Fig4}). Anti-correlated measurements correspond to those where the two projectors at the different telescopes are different, i.e., $\Pi_+$ is measured at one telescope while $\Pi_-$ at the other, in this case the pairs of detectors $L_1,R_2$ or $L_2,R_1$ detect the photons. The probability of anti-correlated photon-detections is
\begin{align} q_{ac}=\Tr[\Pi_+^L\Pi_-^R\rho_A\otimes\rho_X]+\Tr[\Pi_-^L\Pi_+^R\rho_A\otimes\rho_X].
\end{align}
In terms of the matrix elements of the resource state $\rho_X$, given in Eq.~(\ref{xstate}), the probabilities of correlated and anti-correlated measurements turn out to be
\begin{align}
q_{\text{c}} = \frac{1}{4} \left(g+f-2 V_{a} w_{a} \cos (V_{p}-w_{p})\right)\, ,
\label{rawprobs1}
\end{align}
\begin{align}
q_{\text{ac}} = \frac{1}{4} \left(g+f+2 V_{a} w_{a} \cos (V_{p}-w_{p})\right) \, .
\label{rawprobs2}
\end{align}

Postselection on successful detection events, when one detector at each telescope clicks, provides normalized probabilities
\begin{align}
p_{\text{c}} &=\frac{q_c}{q_c+q_{\text{ac}}}= \frac{1}{2} (1- V_{a} C \cos (V_{p}-w_{p})),\\
p_{\text{ac}} &=\frac{q_\text{ac}}{q_c+q_{\text{ac}}}= \frac{1}{2} (1+ V_{a} C \cos (V_{p}-w_{p})) \, ,
\label{postprobs}
\end{align}
where $C=2w_{a}/(g+f)$ is the $X$-state concurrence~\cite{conc-wootters}. The difference between the correlated and anti-correlated event probabilities
\begin{align}
p_{\text{ac}}-p_{\text{c}} =V_a C \cos (V_{p}-w_{p})
\label{visibx}
\end{align}
provides the connection  between the phase, $V_p$, and the amplitude, $V_a$, of the visibility function.
Eq.~(\ref{visibx}) also implies that by controlling and scanning through a few different values of the phase $w_p$ we can determine the values of $V_a$ and $V_p$. Thus, using two values of $w_p$ we have
\begin{align}
\delta p^{(1)}&=p^{(1)}_{ac}-p^{(1)}_c=V_a C \cos (V_{p}-w^{(1)}_{p}) \, , \label{visib-ratio1}\\
\delta p^{(2)}&=p^{(2)}_{ac}-p^{(2)}_c=V_a C \cos (V_{p}-w^{(2)}_{p}) \, . \label{visib-ratio2}
\end{align}
Taking the ratio $\alpha=\delta p^{(1)}/\delta p^{(2)}$, from Eqs.~(\ref{visib-ratio1}), (\ref{visib-ratio2}) we find  the phase of the visibility function
\begin{align}
V_p=&\tan^{-1}[\frac{1}{\sin w^{(2)}_p}(\frac{\sin(w_p^{(2)}-w_p^{(1)})}{\alpha\sin w_p^{(2)} -\sin w_p^{(1)}}-\cos w_p^{(2)})].
\end{align}
Once $V_p$ is determined it can be used to find the magnitude of the visibility function
\begin{align}
V_a&=\frac{\delta p^{(1)}}{C \cos(V_p-w_p^{(1)})} \, .
\label{va}
\end{align}

The rate at which successful detection events occur is called the measurement rate $R_M$. Both correlated and anti-correlated detection events comprise the set of successful detection events. The probability that for a single incoming target photon a successful detection event occurs is $q_M=q_c+q_{\text{ac}}=(g+f)/2$, using Eqs.~(\ref{rawprobs1}), (\ref{rawprobs2}). We denote the photon flux from the target received by the telescopes as $R_T$ and the generation rate of pure Bell-pairs by the photon source in the quantum-network by $R_E$ - which is the rate of generated entangled photons per incoming spatio-temporal mode of astronomical photons with $0\leq R_E\leq 1$. The measurement rate is a $q_M$ fraction of the rate $R_E R_T$ yielding,
\begin{align}
R_M=\frac{(g+f)}{2}R_E R_T \, .
\end{align}
If the entangled states generated at the source suffer no decoherence on their way to the telescopes then $g+f=1$ and the maximal measurement rate for fixed $R_E$ and $R_T$ is $R_M^{(0)}=R_E R_T/2$. A useful way to express the measurement rate $R_M$ is as the product, $R_M=\xi R_M^{(0)}$, with $\xi=g+f$ extracting the contribution to the rate from the matrix elements of the resource state $\bm \rho_X$. The factor of $1/2$ reduction in measurement rate, even when $\xi=1$ and $R_E=1$, arises from the fact that for the ideal case of Bell-state resource both photons end up at the detectors of the same telescope 50\% of the time. A way to reduce this inherent fraction of loss is suggested in~\cite{prl-gottesman-lb} based on the idea of using multipartite entangled resource states distributed between multiple telescopes instead of the bipartite entangled state in the current discussion.

\subsection{Error in visibility measurement due to imperfectly entangled resource states}
\label{subsec:error}

Now we examine the dependence of the errors in the measurement of  visibility function on the parameters  of the resource state. The phase $V_p$ depends on the ratio $\alpha=\delta p^{(1)}/\delta p^{(2)}$ and the phases of the resource state density matrix element $w_p^{(1),(2)}$. The errors in $V_p$ depend only on the errors in the ratio $\alpha$, i.e., $\Delta V_p=|\partial V_p/\partial \alpha|\Delta \alpha$ since we assume the ability to characterize the resource state perfectly and hence $w_p$ is known exactly. For the error in $\alpha$, we have
\begin{align}
\Delta \alpha=\left\{\frac{\partial \alpha}{\partial (\delta p^{(1)})}\Delta(\delta p^{(1)})^2+\frac{\partial \alpha}{\partial (\delta p^{(2)})}\Delta(\delta p^{(2)})^2\right\}^{1/2}.
\end{align}
Since $\Delta(\delta p^{(1)})=\Delta(p^{(1)}_{ac}-p^{(1)}_c)=\Delta(2p^{(1)}_{ac}-1)=2\Delta p^{(1)}_{ac}$ and similarly $\Delta(\delta p^{(2)})=2\Delta p^{(2)}_{ac}$, we use the estimate for the standard statistical error in the determination of the probabilities, i.e., $\Delta p^{(1)}_{ac}=\sigma_{p^{(1)}_{ac}}/\sqrt{N}$ and $\Delta p^{(2)}_{ac}=\sigma_{p^{(2)}_{ac}}/\sqrt{N}$, where $\sigma_{p}$ is the standard deviation of the probability $p$ using a sample of size $N$. Because the number of samples is directly proportional to the measurement rate $R_M$ we find that the error in the measurement of the phase of the visibility, $\Delta V_p$, scales inversely proportional to the square root of the measurement rate, $\Delta V_p\sim 1/\sqrt{N}\sim 1/\sqrt{R_M}\sim1/\sqrt{\xi}$.
The amplitude $V_a$ depends on the phase $V_p$, $\delta p^{(1)}$, the concurrence $C$, and the phase of the resource state density matrix element $w_p^{(1)}$. Our assumption is that $C$ and $w_p^{(1)}$ can  be determined exactly, hence, the errors in $V_a$ are due to errors in the determination of $\delta p^{(1)}$ and the errors in $V_p$. Therefore, we estimate
\begin{align}
\Delta V_a&=\left\{(\frac{\partial V_a}{\partial(\delta p^{(1)})})^2(\Delta \delta p^{(1)})^2+(\frac{\partial V_a}{\partial V_p})^2(\Delta V_p)^2\right\}^{1/2} \, .
\label{va-err}
\end{align}

Using Eq.~(\ref{va}), we find that the partial derivatives in Eq.~(\ref{va-err}) are both proportional to $1/C^2$ and thus $\Delta V_a\propto 1/C$. Further, since both $\Delta \delta p^{(1)}$ and $\Delta V_p$ scale as $1/\sqrt{R_M}$ as discussed in the paragraphs above we find that $\Delta V_a\sim 1/C\sqrt{N}\sim 1/(C\sqrt{R_M})\sim1/(C\sqrt{\xi})$. To summarize, the dependence of the  amplitude and the phase errors in the complex visibility on the parameters of the shared quantum resource (for fixed $R_E$ and $R_T$) goes as follows
\begin{align}
\Delta V_a&\sim1/(C\sqrt{\xi}) \, ,\\
\Delta V_p&\sim1/\sqrt{\xi} \,.
\label{errscaling}
\end{align}

\subsection{Error in the intensity distribution}
\label{subsec:interror}

The errors in the visibility propagate to the errors in the intensity distribution function. The linearity of the Fourier transform, that connects visibility to the intensity, implies that, to first order, the error in the intensity distribution is proportional to $\{\Delta V_a^2+\Delta V_p^2\}^{1/2}$.
Introducing averaged values, $\bar{V}_a$ and $\bar{V}_p$,  respectfully for the amplitude and the phase of the visibility function, so that $V_a=\bar{V}_a+\Delta V_a$ and $V_p=\bar{V}_p+\Delta V_p$, we find
\begin{align}
I(\bm s)&=\int (\bar{V}_a+\Delta V_a)e^{i(\phi(r)+\Delta V_p)}d\bm r \nonumber\\
&\simeq \bar{I}+\int \Delta V_a e^{i\phi(\bm r)}d\bm r + i \Delta V_p \int \bar{V}_a e^{i\phi(\bm r)}  d\bm r\, ,
\end{align}
where $\bar{I}$ is the averaged intensity, and $\phi(\bm r)=2\pi\nu \bm s \bm r/c+\bar{V}_p$.
Therefore, we have
\begin{align}
\Delta I = |I-\bar{I}| &\sim \sqrt{\Delta V_a^2+\Delta V_p^2} \, .
\end{align}

In the estimate above, $\Delta V_a$ and $\Delta V_p$ should be understood as the maximum values of amplitude and phase errors over the domain of integration. From the dependence of the errors on the resource state characteristics given in Eq.~(\ref{errscaling}) we find that the errors in the intensity distribution goes as $(C^2+1)^{1/2}/C\sqrt{\xi}$. Thus, for $C\simeq 1$ the error in the intensity distribution goes as $1/\sqrt{\xi}$, i.e., as the error in the phase of the visibility, while for small values of the concurrence $C$, it goes as $1/C\sqrt{\xi}$, i.e., as the error in the amplitude of the visibility.

\subsection{Examples cases for decoherence of resource states}
\label{subsec:examples}

Here we consider several examples of the decoherence in quantum network components, that result in imperfect resource states for interferometric measurements.

\begin{figure}
\centering
\includegraphics[width=\columnwidth]{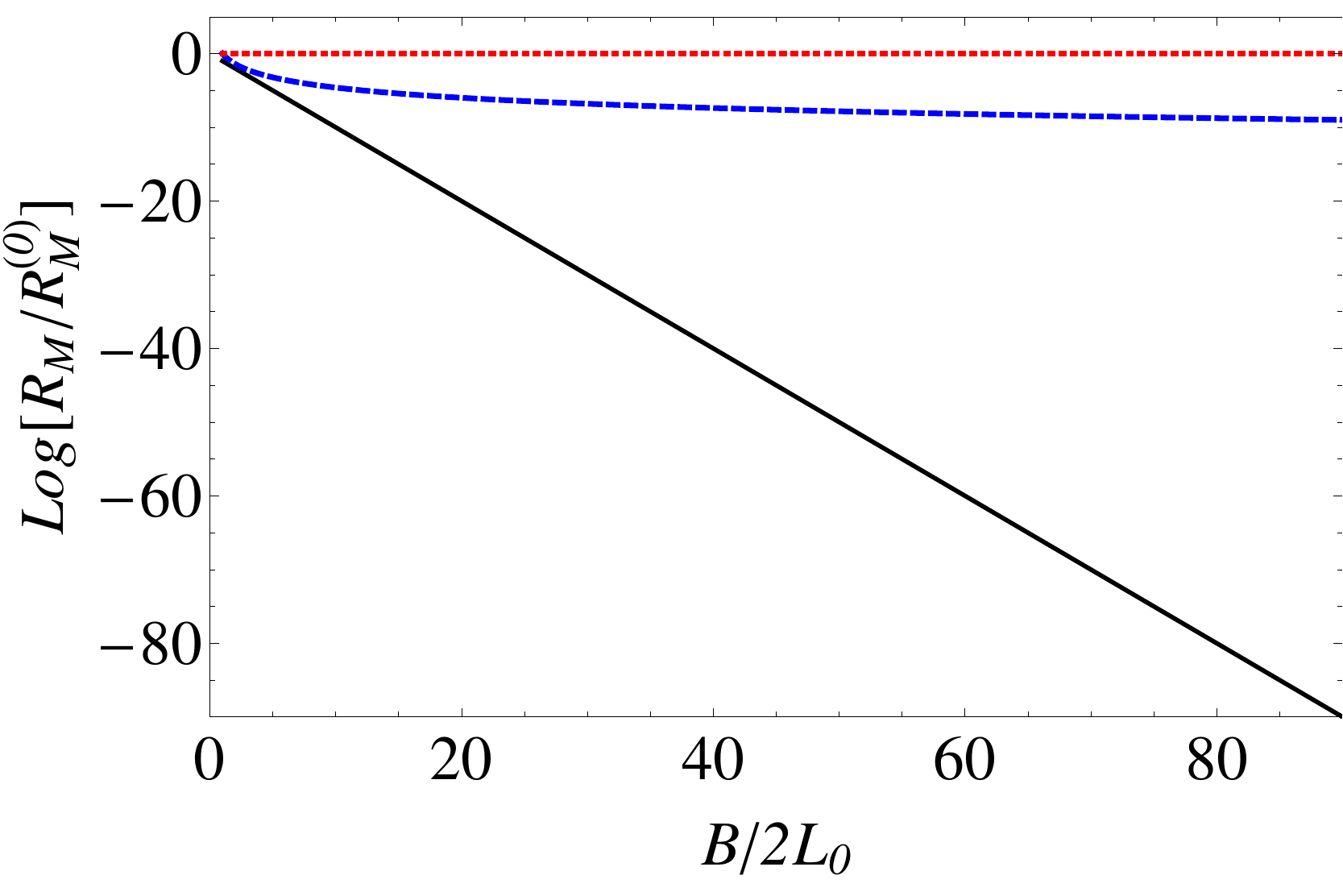}
\caption{(Color online) Normalized measurement rate $R_M$ as a function of normalized baseline $B$ for three cases: channel with exponential loss (solid black line), quantum repeater chain with polynomial loss (dashed blue line), perfect repeater chain distributing ideal Bell pairs (dotted red line).}
\label{Fig5}
\end{figure}

\emph{Lossy Fibers in quantum network.} If the entangled photon in the quantum-network is distributed to the telescopes using lossy optical fibers, the photon loss can be modeled as amplitude-damping decoherence. Considering the decoherence only due to amplitude-damping, from Eq.~(\ref{ampstate}) we get $g+f=1-(\lambda_L+\lambda_R)/2$, where $\lambda_{L,R}=(1-e^{-L_{L,R}/L_0})$ is the probability of photon loss in the fiber of a length $L$  with $L_0$ being the attenuation length. In the case of of equal arm lengths of the two channels, $L_L=L_R=B/2$, shown in Fig.~(\ref{Fig4}), we find that the log of the measurement rate, shown in Fig.~(\ref{Fig5}), goes down linearly with the size of the baseline
\begin{align}
\log(R_M)&= \log(R_E R_T/2)-\frac{B}{2L_0} \, .
\label{targetrate2}
\end{align}
With equal arm lengths we also observe that the errors in the amplitude and phase of the visibility function  depend on the baseline size as $\Delta V_a\sim e^{B/4L_0}$ and $\Delta V_p\sim e^{B/4L_0}$.

\begin{figure}
\centering
\includegraphics[width=\columnwidth]{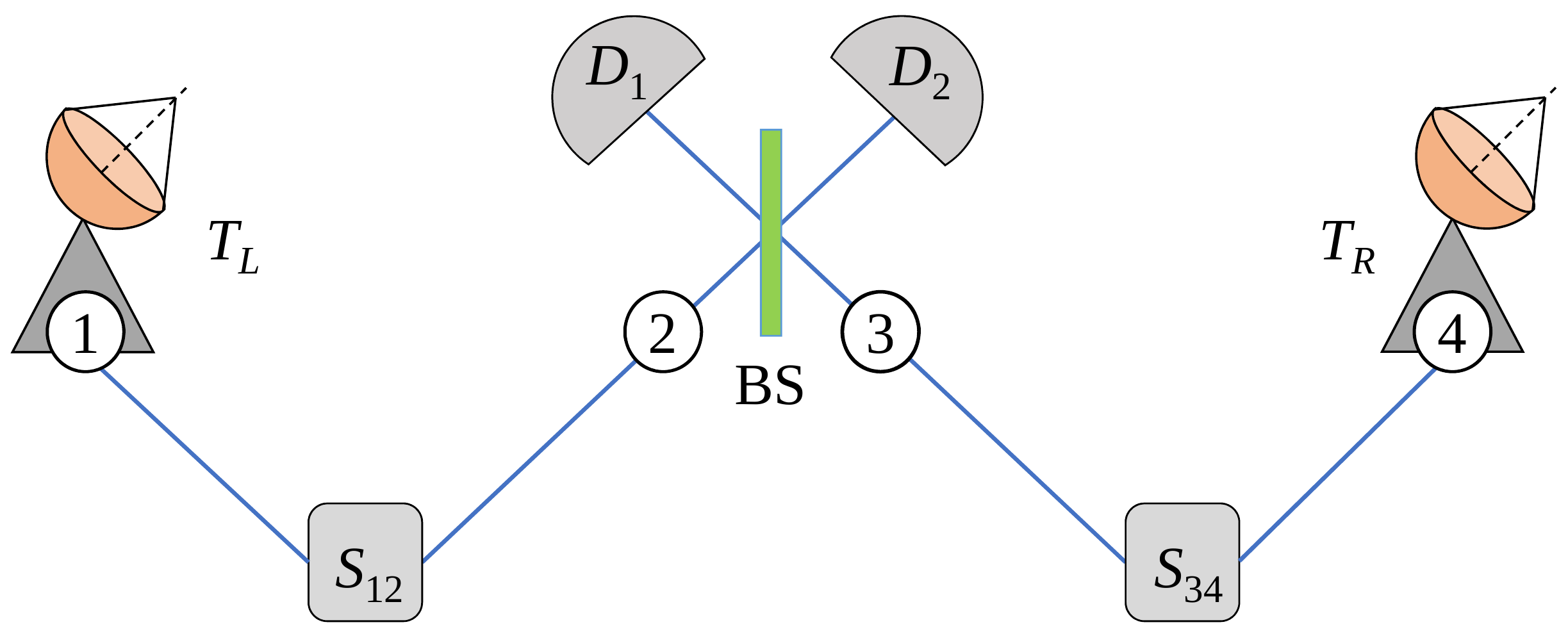}
\caption{(Color online) Schematic of a quantum-network based on entanglement swapping of states stored in quantum memories. $S_{12},S_{34}$ are the sources of entangled Bell-pairs of photons. The Bell-pair produced by $S_{12}$ is stored in quantum memories $1,2$ while that produced by $S_{34}$ in $3,4$. Joint measurements on the quantum state in memories $2$ and $3$ using a linear optical setup such as a beam splitter (BS) and photon detectors $D_1$, $D_2$ result in an entangled quantum state of memories $1$ and $4$. This produces the resource quantum-state shared between the telescopes $T_L$ and $T_R$.}
\label{Fig6}
\end{figure}

\emph{Finite-lifetime quantum memories.} Another example of decoherence in the network that leads to an X-state resource is dephasing in quantum memories. To show this, we consider a quantum network which relies on quantum memories and entanglement swapping to distribute the state between the two telescopes as shown in Fig.~(\ref{Fig6}). The simplest scheme comprises of two sources of entangled Bell-pairs of photons, a set of quantum memories at the telescope sites, and an entanglement swapping setup at the middle station. One photon of each entangled pair is stored in a quantum memory at the telescope site before entanglement swapping is performed by a joint measurement on one photon from each pair at a central station~\cite{duan2001long,razavi}. The action of an imperfect memory on a qubit $\sigma$ stored in the memory for a time $t$ can be modeled as a dephasing. Following~\cite{razavi}, we describe  a single-qubit dephasing using a superoperator, $\hat{\bm \Gamma}_t$, acting on the qubit density matrix, $\bm \sigma$, as
\begin{equation}
\hat{\bm \Gamma}_{t} \bm \sigma=p(t/2)\bm \sigma+[1-p(t/2)]\bm Z \bm \sigma \bm Z,
\end{equation}
where $p(t)=(1+e^{-t/\tau_{c}})/2$, $\tau_{c}$ is the memory coherence time, $\bm Z$ is the spin Pauli operator along the $z$-direction.

When each qubit of an initial entangled Bell-state,  $\psi_\pm$,  stored in identical quantum memories with lifetime $\tau_c$, the decohered state at time $t$ due to dephasing can be described as
\begin{equation}
\hat{\bm \Gamma}_{t} \otimes \hat{\bm \Gamma}_{t} \bm \rho^{\psi \pm}=p(t) \bm \rho^{\psi \pm}+[1-p(t)] \bm \rho^{\psi \mp}=\bm \rho_{M}^{\pm}(t),
\end{equation}
where $\bm \rho^{\psi \pm}$ is the density matrix of the original Bell state. After some algebra we find
\begin{equation}
\bm \rho_{M}^{\pm}(t)=
\left(
\begin{array}{cccc}
 0 & 0 & 0 & 0 \\
 0 & \frac{1}{2} & \pm\frac{1}{2} e^{-\frac{t}{\tau_c }} & 0 \\
 0 & \pm\frac{1}{2} e^{-\frac{t}{\tau_c }} & \frac{1}{2} & 0 \\
 0 & 0 & 0 & 0 \\
\end{array}
\right) \, ,
\end{equation}
which is the density matrix of the Bell diagonal state.

Entanglement swapping of two states $\bm \rho_{M}^{\pm}(t_{1})$ and $\bm \rho_{M}^{\pm}(t_{2})$  stored in the different quantum memories for the time $t_{1}$ and $t_{2}$ results in the final output state described by
\begin{equation}
\bm \rho_{M}^{\pm}(t_{1}+t_{2})=p(t_{1}+t_{2}) \bm \rho_{M}^{\psi \pm}+[1-p(t_{1}+t_{2})] \bm \rho_{M}^{\psi \mp},
\end{equation}
where the $\pm$ superscript  denotes whether the Bell-state measurement outcome for the entanglement-swapping procedure is the $\psi^{\pm}$ Bell state. This result is valid for both $\psi^{+}$ and $\psi^{-}$ Bell states as the input states. Comparing the density matrix $\bm \rho_{M}^{\pm}(t_{1}+t_{2})$ to $\bm \rho_X$ in Eq.~(\ref{xstate}) we find that $g+f=1$,  thus the measurement rate $R_M$ remains constant. This  implies that the error in the phase of the visibility function is also independent of the times $t_1$ and $t_2$. The error in the amplitude of the visibility goes as $\Delta V_a\sim e^{(t_1+t_2)/\tau_c}$.

\emph{Birefringent optical fibers.} Finally, we consider decoherence due to depolarization. Polarization-entangled photons can also be used as a resource to increase the telescope baseline. Control and manipulation of the polarization can provide a way to obtain single photon spatial-mode entangled photons that match the spatio-temporal mode of the incoming astronomical photons. From the density matrix of the $X$-state, Eq.~(\ref{depolstate}), it is clear that the measurement rate in this case is $R_M=(1-2x)R_E R_T/2$ with $x=(\kappa_L+\kappa_R)/3-4\kappa_L\kappa_R/9$. Describing the depolarization probabilities in the left and right network channels as $\kappa_L=\kappa_R=\kappa=(1-e^{-\beta L/2})$, where $\beta$ is the inverse attenuation length for depolarization and considering large channel lengths, such that $e^{-\beta L}\ll 1$, we find that the logarithm of the measurement rate is well approximated by
\begin{align}
\log(R_M)\approx \log(R_E R_T/2)+\log(5/9)+0.8 e^{-\beta L/2} \, .
\end{align}
The measurement rate in this case converges to a constant $R_M\simeq 0.28 R_E R_T$ which follows from the fact that the matrix elements $g$ and $f$ do not decay to zero for any length of the fiber, see Eq.~(\ref{depolstate}).

\section{Conclusions}
\label{sec:conc}

In this work we demonstrated that even imperfectly entangled quantum states are a useful resource to increase the telescopic baseline size in direct-detection interferometry. We showed the dependence of the visibility function, the measurement rate, and the errors in the amplitude and phase of the visibility on the the density matrix parameters of the entangled $X$-state. As a resource for visibility measurement, we found that distributed entangled quantum-states can be characterized by their concurrence, $C$, and the weight, $\xi$, of their density matrix in a relevant mode-entangled subspace. We showed that the measurement rate is proportional to $\xi$ of the resource state. Focusing on the entangled $X$-states resulting from the perfect Bell-pairs due to decoherence, we performed error analysis of the visibility function measurement scheme. Several physical examples - amplitude-damping due to lossy-fibers, dephasing in quantum memories, and depolarization of polarization-entangled photons were considered. We found that the error in the amplitude of the visibility, $V_A$, is inversely proportional to the product of the concurrence and square root of the weight of the resource state, i.e, $\Delta V_a\sim1/C\sqrt{\xi}$. The error in the phase of the visibility $V_p$ is inversely proportional to only the square root of the weight, i.e., $\Delta V_p\sim1/\sqrt{\xi}$. Further, we derived explicit formulae for the dependence of the resource-state characteristics $C$ and $\xi$ on the decoherence parameters of the quantum-network. We conclude that even substantially decohered entangled distributed quantum-states can serve as a robust resource for improved measurements of interferometric visibility.

The scheme, even with imperfect resources, can mitigate photon loss and provide better resolution when compared to other interferometric measurement methods~\cite{review-monnier}, such as intensity interferometry based on the Hanbury Brown Twiss effect~\cite{hanburytwiss} and heterodyne interferometry.
It yields both the amplitude and phase of the visibility function whereas intensity interferometry usually loses the phase information. With respect to heterodyne interferometry the quantum enhanced direct-detection scheme can have a better signal-to-noise ratio in the optical regime where the former is limited by quantum noise. The entanglement of the resource-state correlates the noise in the measurement outcomes at different telescopes which gets cancelled by considering their correlation or anti-correlation. Implementation of the generalized quantum-enhanced direct-detection interferometry would require high throughput quantum-networks that can generate mode-entangled single photons over a wide range of wavelengths to maximize the fraction of incoming photons succesfully detected. To estimate required entanglement generation rates we consider a reference star. For instance, a star called `Vega' produces a photon flux of a few MHz for a telescope of diameter $1$m. Entanglement generation rates of MHz or above should be feasible in the near future.

\emph{Acknowledgement.}
The authors thank Alejandra Maldonado-Trapp, Sreraman Muralidharan, Linshu Li, Liang Jiang and Charles Clark for useful conversations and correspondence. This work was supported in part by the Office of the Secretary of Defense, Quantum Science and Engineering Program.

\appendix
\section{Decoherence in the quantum channel}
\label{app:xform}
\emph{Amplitude Damping decoherence.} Independent amplitude damping in the left ($L$) and right ($R$) arms  of the fiber-optic link can be described by the Kraus operators
\begin{align}
 A^{L,R}_1&=\ket{0}\bra{0}+\sqrt{1-\lambda_{L,R}}\ket{1}\bra{1}\nonumber\\ A^{L,R}_2&=\sqrt{\lambda_{L,R}}\ket{0}\bra{1} \, ,
\end{align}
with $\sum_{i=1,2}(A^{L,R})^\dagger_iA^{L,R}_i=\openone_{2}$, and $\lambda_{L,R}$ is the probability of the photon loss in the left and the right arm of the channel. Thus, joint decoherence in the quantum link can be described by the tensor products of the left and right arm operators, $A^L_1\otimes A^R_1,A^L_1\otimes A^R_2,A^L_2\otimes A^R_1,A^L_2\otimes A^R_2$. The final 2-qubit state emerging through the quantum channel for the initial state $\bm \rho_i$ is given by $\bm \rho_o=\sum_{i,j=1,2}A^L_i\otimes A^R_j~\bm \rho_i~(A^L)^\dagger_i\otimes (A^R)^\dagger_j$. Thus for $\bm \rho_i=\ket{\psi_+}\bra{\psi_+}$, where $\ket{\psi_+}=(\ket{0}\ket{1}+\ket{1}\ket{0})/\sqrt{2}$, we have
\begin{equation}
\bm \rho_{X}=\left(
\begin{array}{cccc}
 \frac{\lambda_L+\lambda_R}{2} & 0 & 0 & 0 \\
 0 & \frac{1-\lambda_R}{2} & \frac{\sqrt{(1-\lambda_L)(1-\lambda_R)}}{2} & 0 \\
 0 & \frac{\sqrt{(1-\lambda_L)(1-\lambda_R)}}{2} & \frac{1-\lambda_L}{2} & 0 \\
0 & 0 & 0 & 0 \\
\end{array}
\right) \, .
\label{ampstate1}
\end{equation}

\emph{Dephasing decoherence.} Independent dephasing effects in the two network arms can be represented by Kraus operators
\begin{align}
P^{L,R}_1&=\sqrt{1-\mu_{L,R}}\openone_2 \, , \nonumber\\
P^{L,R}_2&=\sqrt{\mu_{L,R}}\ket{0}\bra{0} \, , \nonumber\\
P^{L,R}_3&=\sqrt{\mu_{L,R}}\ket{1}\bra{1} \, ,
\end{align}
which satisfy the quantum operation condition $\sum_{i=1,2,3}(P^{L,R})^\dagger_iP^{L,R}_i=\openone_{2}$. The joint dephasing is then described by operators of the form $P^L_i\otimes P^R_j,~i,j=\{1,2,3\}$ which result in a final 2-qubit state $\bm \rho_o=\sum_{i,j=1,2,3}P^L_i\otimes P^R_j~\bm \rho_i~(P^L)^\dagger_i\otimes (P^R)^\dagger_j$. Again for the case of the initial state $\bm \rho_i=\ket{\psi_+}\bra{\psi_+}$ we have
\begin{equation}
\rho_{X}=\left(
\begin{array}{cccc}
0 & 0 & 0 & 0 \\
 0 & .5 & .5(1-\mu_L)(1-\mu_R) & 0 \\
 0 & .5(1-\mu_L)(1-\mu_R) & .5 & 0 \\
0 & 0 & 0 & 0 \\
\end{array}
\right) \, .
\label{dephstate1}
\end{equation}

\emph{Depolarizing decoherence.} Independent depolarization effects in the network arms can be described by operators
\begin{align}
M^{L,R}_1&=\sqrt{(1-\kappa_{L,R})}~\openone_2 \, , \nonumber\\
M^{L,R}_2&=\sqrt{\kappa_{L,R}/3}~\sigma_x \, , \nonumber\\
M^{L,R}_3&=\sqrt{\kappa_{L,R}/3}~\sigma_y \, , \nonumber\\
M^{L,R}_4&=\sqrt{\kappa_{L,R}/3}~\sigma_z \, ,
\end{align}
which satisfy $\sum_{i=1,2,3,4}(M^{L,R})^\dagger_iM^{L,R}_i=\openone_{2}$. Joint depolarization is thus given by operators $M^L_i\otimes M^R_j,~i,j=\{1,2,3,4\}$ and the output state $\bm \rho_o=\sum_{i,j=1,2,3,4}M^L_i\otimes M^R_j~\bm \rho_i~(M^L)^\dagger_i\otimes (M^R)^\dagger_j$ which for $\bm \rho_i=\ket{\psi_+}\bra{\psi_+}$ is given by
\begin{equation}
\rho_X=\left(
\begin{array}{cccc}
 x & 0 & 0 & 0 \\
 0 & .5-x & .5-2x & 0 \\
 0 & .5-2x & .5-x & 0 \\
 0  & 0 & 0 & x \\
\end{array}
\right) \, ,
\label{depolstate1}
\end{equation}
where $0\leq x=(\kappa_L+\kappa_R)/3- 4\kappa_L\kappa_R/9 \leq 1/3$, since $0\leq \kappa_{L,R} \leq 1$.

It is clear from the form of the states Eq.~(\ref{ampstate}), (\ref{dephstate}), (\ref{depolstate}) that independent amplitude-damping, dephasing and depolarization in the  quantum channel arms between the source and the telescopes all lead to states of the X-form. Indeed, because this form of the resultant state is independent of the type of decoherence in the two arms, one can imagine different combinations of decoherence mechanisms, e.g. $A^L\otimes P^R, M^L\otimes A^R$ etc., which preserve the X-form.\\

\section{Errors in the visibility function for various decoherence mechanisms}
Various decoherence mechanisms outlined in Sec.~(\ref{subsec:decoh}) result in imperfect resource states and errors in the visibility measurement. The visibility errors due to the various decoherence types  depend on the parameters of the resource states scale as
\begin{align}
\text{Amplitude Damping:} & \Delta V_a\sim \frac{(1-.5(\lambda_L+\lambda_R))^{1/2}}{R_X^{1/2} (1-\lambda_L)^{1/2}(1-\lambda_R)^{1/2}} \, , \nonumber\\
&\Delta V_p\sim \frac{1}{R_X^{1/2}(1-.5(\lambda_L+\lambda_R))^{1/2}} \, , \nonumber\\
\text{Dephasing:} & \Delta V_a\sim \frac{1}{R_X^{1/2} (1-\mu_L)(1-\mu_R)} \, , \nonumber\\
&\Delta V_p\sim \frac{1}{R_X^{1/2}} \, , \nonumber\\
\text{Depolarization:}& \Delta V_a\sim \frac{(1-2x)^{1/2}}{R_E^{1/2} (1-4x)} \, , \nonumber\\
&\Delta V_p\sim \frac{1}{R_X^{1/2} (1-2x)^{1/2}} \, , \nonumber
\end{align}
where $x=(\kappa_L+\kappa_R)/3-4\kappa_L\kappa_R/9$, and $R_X$ is the rate of network-supplied entangled photons per incoming spatio-temporal mode of astronomical photons.

\bibliography{refs-qtelescope}
\bibliographystyle{apsrev4-1}

\end{document}